\documentclass[pra,aps,preprintnumbers,showpacs,superscriptaddress]{revtex4}
\usepackage{epsfig}
\usepackage{graphics}

\newcommand{\qed}{\nobreak \ifvmode \relax \else
      \ifdim\lastskip<1.5em \hskip-\lastskip
      \hskip1.5em plus0em minus0.5em \fi \nobreak
      \vrule height0.75em width0.5em depth0.25em\fi}
\newcommand{\beq}{\begin{equation}}
\newcommand{\eeq}{\end{equation}}

\def\opone{\leavevmode\hbox{\small1\kern-3.8pt\normalsize1}}

\input{tcilatex}
\begin{document}

\title{ Wigner-Yanase-Dyson information as a measure of quantum uncertainty
of mixed states }
\author{Dafa Li$^{a}$\thanks{%
email address:dli@math.tsinghua.edu.cn}}
\affiliation{$^a$ Dept of Mathematical Sciences, Tsinghua University, Beijing 100084 China}
\author{Xinxin Li}
\affiliation{Dept. of Computer Science, Wayne State University, Detroit, MI 48202, USA}
\author{Hongtao Huang}
\affiliation{Electrical Engineering and Computer Science Department, University of
Michigan, Ann Arbor, MI 48109, USA }
\author{ Xiangrong Li}
\affiliation{Department of Mathematics, University of California, Irvine, CA 92697-3875}
\author{L. C. Kwek}
\affiliation{National Institute of Education, Nanyang Technological University, 1 Nanyang
Walk, Singapore 637616}
\affiliation{Centre for Quantum Technologies, National Univeristy of Singapore, 3 Science
Drive 2, Singapore 117543 }
\affiliation{Institute of Advanced Studies (IAS), Nanyang Technological University, 60
Nanyang View Singapore 639673}

\begin{abstract}
In this paper, we consider Wigner-Yanase-Dyson information as a measure of
quantum uncertainty of a mixed state. We study some of the interesting
properties of this generalized measure. The construction is reminiscent of
the generalized entropies that have shown to be useful in many applications.
\end{abstract}

\maketitle

.

\section{Introduction}

Entropy is a measure of the lack of information about a system \cite{wehrl}.
It can also be regarded as the amount of uncertainty in the outcomes of a
measurement on a system. In information theory, Shannon developed
information entropy as a measure of uncertainty in a message\cite{shannon}.
This entropy was generalized in the quantum context to von Neumann entropy
which is defined for a mixed state $\rho $ as $S(\rho )=-\mbox{\rm Tr}\rho
\log \rho $. Let $\{\lambda _{i}\}$ be the spectrum of$\ $the state $\rho $.
Then von Neumann entropy of $\rho $ can be rewritten as $S(\rho )=-\sum
\lambda _{i}\log \lambda _{i}$, where $0\log 0=0$. For example, for an $n$%
-dimensional maximally mixed state $\rho =I/n$, the direct computation gives
$S(\rho )=\log n$. Also see \cite{Nielsen}. For a pure state, $\psi \rangle $%
, $S(\psi \rangle \langle \psi |)=0$. Whereas for a maximally mixed state,
it acquires its maximal value of $\log n$, where $n$ is the dimension of the
density matrix $\rho $.

Indeed, it is well known by now that von Neumann entropy, which is based on
Shannon entropy for an information system, is a unique measure that
satisfies the four Khinchin axioms \cite{khinchin}. Two of the axioms are
convexity and additivity. Relaxing the convexity requirement leads to Renyi
entropy defined by $\displaystyle S^{R}(\rho )=\frac{\log \mbox{\rm Tr}\rho
^{q}}{q-1},$ while relaxing the additivity condition gives Tsallis entropy $%
\displaystyle S^{T}(\rho )=\frac{1-\mbox{\rm Tr}\rho ^{q}}{q-1},$ where $q$
is some adjustable parameter. In both cases, one recovers von Neumann
entropy in the limit $q\rightarrow 1$. These generalized entropies have
found applications in a wide variety of situations: Renyi entropy has been
useful for the analysis of channel capacities \cite%
{renner,giovannetti,birula} and Tsallis entropies have been applied
successfully to some physical situations like multiparticle processes in
particle physics\cite{wilk1,wilk2}. In \cite{Hu}, some of the generalized
quantum entropies were introduced, and nonnegativity, continuity and
concavity were discussed. However, the additivity and subadditivity do not
always hold for these entropies \cite{Hu}.

However, it is argued that the quantum uncertainty of $\rho =I/n$\ should
vanish \cite{Luo2, Brukner}. Brukner and Zeilinger discussed conceptual
inadequacy of the Shannon information in quantum measurement \cite{Brukner}.
They suggested a new measure of information for an individual measurement
with $n$ possible outcomes, and the measurement of the total information $%
I_{total}=\mbox{\rm Tr}\rho ^{2}-1/n$, where $\rho $ is the density
operator. Moreover, since von Neumann entropy vanishes for all pure states,
Wigner and Yanase proposed an entropy which measures our knowledge of a
difficult-to-measure observable with respect to a conserved quantity. They
defined the entropy as $I(\rho $, $A)=$ $-\frac{1}{2}\mbox{\rm Tr}\rho
^{1/2} $, $A]^{2}$, relative to a self-adjoint \textquotedblleft
observable", $A$, which they called the skew information\cite{Wigner}.
Recently, the skew information $I(\rho $, $A)=$ $-\frac{1}{2}\mbox{\rm Tr}%
\rho ^{1/2}$, $A]^{2}$ were studied in \cite{Luo1}\cite{Luo2}\cite{Luo3}.%
\cite{Luo4}\cite{Chen}\cite{Hansen}. It was indicated that the skew
information is a kind of Fisher information \cite{Luo3}. Recently, Hansen
demonstrated that the skew information is not subadditive by giving a
counter example \cite{Hansen}. Dyson generalized the skew information as $%
I_{\alpha }(\rho ,X)=-\frac{1}{2}\mbox{\rm Tr}([\rho ^{\alpha }$, $X][\rho
^{1-\alpha }$, $X])$, usually called the Wigner-Yanase-Dyson entropy, where $%
0<\alpha <1$. Ref. \cite{Wigner}\cite{Lieb}. When $\alpha =1/2$, it reduces
to the skew information. Hansen also reported that the Wigner-Yanase-Dyson
entropy is not subadditive \cite{Hansen}. Uncertainty principles for
Wigner-Yanase-Dyson information were investigated in \cite{Paolo}\cite{dli}%
.\ By calculating, $I_{\alpha }(\rho ,X)$ can be rewritten as
\begin{equation}
I_{\alpha }(\rho ,X)=\mbox{\rm Tr}(\rho X^{2})-\mbox{\rm Tr}(\rho ^{\alpha
}X\rho ^{1-\alpha }X).  \label{Dyson-1}
\end{equation}%
\

It is well known that the following variance of the observable $X$ in the
quantum state $\rho $
\begin{equation}
V(\rho ,X)=\mbox{\rm Tr}(\rho X^{2})-(\mbox{\rm Tr}(\rho X))^{2}
\label{variance}
\end{equation}%
is a primary uncertainty measure. The variance depends on the observable $X$
and includes quantum and classical uncertainty. To be rid of the observable $%
X$, it is intuitive to average the variance over the observables. Instead of
averaging the variance, Luo averaged the skew information \cite{Luo2}. In
\cite{Luo2}, he defined the quantum uncertainty for a mixed state $\rho $ of
an $n$-dimensional quantum system\ as $L(\rho )=\sum_{j=1}^{n^{2}}I(\rho
,H_{j})$ over an orthonormal basis $\{H_{j}\}$ for the real $n^{2}$
dimensional Hilbert space of the observables with inner product $\langle
X,Y\rangle =\mbox{\rm Tr}(XY)$, and demonstrated that the quantity $L(\rho )$
is independent on the choice of the orthonormal basis. By using the property
$I(U\rho U^{\dagger }$, $H)=I(\rho ,UHU^{\dagger })$\cite{Luo2}, Luo showed
that $L(\rho )$ is invariant under unitary transformations, i.e., $L(U\rho
U^{\dagger })=L(\rho )$. It is well known that for some unitary $U$, $U\rho
U^{\dagger }=diag\{\lambda _{1}$, $\lambda _{2}$,..., $\lambda _{n}\}$,
where $\{\lambda _{i}\}$ is the spectrum of $\rho $. Thus, without loss of
the generality, it can be assumed that $\rho =D=diag\{\lambda _{1}$, $%
\lambda _{2}$,..., $\lambda _{n}\}$. Then for any observable $H$, the
straightforward calculation of $I(D,H)$ yields%
\begin{equation}
I(D,H)=\sum_{i<k}(\sqrt{\lambda _{i}}-\sqrt{\lambda _{k}})^{2}\left\vert
\left\vert h_{ik}\right\vert \right\vert ^{2},  \label{Luo-0}
\end{equation}%
where $h_{ik}$ is the entry $(i,k)$ of $H$. By choosing the special
orthonormal basis \cite{Luo2}, Luo obtained \cite{Luo2}

\begin{equation}
L(\rho )=L(D)=\sum_{i<k}(\sqrt{\lambda _{i}}-\sqrt{\lambda _{k}})^{2}=n-(%
\mbox{\rm Tr} \sqrt{\rho })^{2},  \label{Luo-1}
\end{equation}%
which is rid of the observables.

\section{Properties of Wigner-Yanase-Dyson (WYD) information}

The WYD information possesses some interesting properties which we will
summarize in this section.

\begin{enumerate}
\item Wigner-Yanase-Dyson information is convex with respect to $\rho $ \cite%
{Lieb}. However, $\mbox{\rm Tr} (\rho ^{\alpha }X\rho ^{1-\alpha }X)$ with
respect to $\rho $\ is concave \cite{Lieb}.

\item Let $\rho _{1}$ and $\rho _{2}$ be two density operators of two
subsystems and let $A_{1}$ (resp. $A_{2}$) be a self-adjoint operator on $%
H^{1}$ (resp. $H^{2}$). Then WYD information $I_{\alpha }(\rho ,X)$
satisfies $I_{\alpha }(\rho _{1}\otimes \rho _{2}$, $A_{1}\otimes
I_{2}+I_{1}\otimes A_{2})=I_{\alpha }(\rho _{1}$, $A_{1})+I_{\alpha }(\rho
_{2}$, $A_{2})$, where $I_{1}$ and $I_{2}$ are the identity operators for
the first and second systems, respectively. See \cite{Lieb}\cite{dli}. The
case in which $\alpha =1/2$ was discussed in \cite{Luo3}.

\item $I_{\alpha }(\rho ,A_{1}\otimes I_{2})\geq I_{\alpha }(\rho
_{1},A_{1}) $, where $\rho _{1}=tr_{2}\rho $. We can argue this as follows.
A simple calculation shows $\mbox{\rm Tr} (\rho (A_{1}\otimes I_{2})^{2})=%
\mbox{\rm
Tr} (\rho _{1}A_{1}^{2})$. By (2.2) in \cite{Lieb}, $\mbox{\rm Tr} (\rho
^{\alpha }(A_{1}\otimes I_{2})\rho ^{1-\alpha }(A_{1}\otimes I_{2}))\leq $ $%
\mbox{\rm Tr} (\rho ^{\alpha }A_{1}\rho ^{1-\alpha }A_{1})$. By the
definition in Eq. (\ref{Dyson-1}), this property holds.

\item When $\rho $ is pure, $V(\rho ,X)=I_{\alpha }(\rho $, $X)$. Thus, the
Wigner-Yanase-Dyson information reduces to the variance. The case in which $%
\alpha =1/2$ was discussed in \cite{Luo1}.

\item When $\rho $ is a mixed state, $V(\rho ,X)\geq I_{\alpha }(\rho $, $X)$%
. This is because $\mbox{\rm Tr} (\rho ^{\alpha }X\rho ^{1-\alpha }X)\geq 0$%
. The case in which $\alpha =1/2$ was discussed in \cite{Luo1}. Also\ see
\cite{dli}.

\item When $\rho $ and $A$ commute, by the discussion in \cite{Luo4} the
quantum uncertainty based on the skew information should vanish.\ It is easy
to verify that Wigner-Yanase-Dyson information $I_{\alpha }(\rho ,X)$ also
satisfies this requirement. We can argue this property from that $\rho $ and
$A$ share an orthonormal eigenvector basis when $\rho $ and $A$ commute \cite%
{Mika}.\

\item The invariance of Wigner-Yanase-Dyson information $I_{\alpha }(\rho
,X) $ under unitary transformations. The case in which $\alpha =1/2$ was
discussed in \cite{Luo2}\cite{Luo4}.

\begin{itemize}
\item $I_{\alpha }(U\rho U^{\dagger },X)=I_{\alpha }(\rho ,U^{\dagger }XU)$
for any unitary operator $U$. See Appendix A.

\item $I_{\alpha }(U\rho U^{\dagger },UXU^{\dagger })=I_{\alpha }(\rho $, $%
X) $ for any unitary operator $U$. See Appendix A.

\item $I_{\alpha }(U\rho U^{\dagger },X)=I_{\alpha }(\rho $, $X)$ for any
unitary operator $U$ if the unitary operator $U$ commutes with $X$.
\end{itemize}
\end{enumerate}

\section{Average Wigner-Yanase-Dyson information as quantum uncertainty}

Rather than averaging the skew information, we propose to average WYD
information. To this end, we propose $Q_{\alpha }(\rho
)=\sum_{j=1}^{n^{2}}I_{\alpha }(\rho ,H_{j})$ as the quantum uncertainty of
a mixed state $\rho $, where $\{H_{j}\}$ is defined as above.\ As discussed
in \cite{Luo2}, we can also show that the quantity $Q_{\alpha }(\rho )$ does
not depend on the choice of the orthonormal basis. Let $\{\lambda _{i}\}$ be
the spectrum of $\rho $. By only means of the spectral representation of $%
\rho $ and the definition of $I_{\alpha }(\rho ,H)$\ in Eq. (\ref{Dyson-1}),
the direct calculation of $I_{\alpha }(\rho ,H)$ for any observable $H$
shows $I_{\alpha }(\rho ,H)=\sum_{i<j}(\lambda _{i}+\lambda _{j}-\lambda
_{i}^{\alpha }\lambda _{j}^{1-\alpha }-\lambda _{i}^{1-\alpha }\lambda
_{j}^{\alpha })\left\vert \left\vert h_{ij}\right\vert \right\vert ^{2}$
\cite{dli}. By choosing the special orthonormal basis in \cite{Luo2}, we
obtain $Q_{\alpha }(\rho )=\sum_{i<j}(\lambda _{i}+\lambda _{j}-\lambda
_{i}^{\alpha }\lambda _{j}^{1-\alpha }-\lambda _{i}^{1-\alpha }\lambda
_{j}^{\alpha })$, which depends only on the mixed state $\rho $.
Furthermore, we rewrite%
\begin{equation}
Q_{\alpha }(\rho )=\sum_{i<j}(\lambda _{i}^{\alpha }-\lambda _{j}^{\alpha
})(\lambda _{i}^{1-\alpha }-\lambda _{j}^{1-\alpha })=n-\mbox{\rm Tr} \rho
^{\alpha }\mbox{\rm Tr} \rho ^{1-\alpha }.  \label{new-1}
\end{equation}%
To demonstrate that $Q_{\alpha }(\rho )$ is less than $n-1$, we rephrase
\begin{equation}
Q_{\alpha }(\rho )=n-1-\sum_{i<k}(\lambda _{i}^{\alpha }\lambda
_{k}^{1-\alpha }+\lambda _{i}^{1-\alpha }\lambda _{k}^{\alpha }).
\label{def-2}
\end{equation}%
This equality follows Eq. (\ref{new-1})\ and $\mbox{\rm Tr} \rho ^{\alpha }%
\mbox{\rm Tr} \rho ^{1-\alpha }=\sum_{i}\lambda _{i}^{\alpha
}\sum_{k}\lambda _{k}^{1-\alpha }=1+\sum_{1\leq i<k\leq n}(\lambda
_{i}^{\alpha }\lambda _{k}^{1-\alpha }+\lambda _{i}^{1-\alpha }\lambda
_{k}^{\alpha })$. When $\alpha =1/2$, $Q_{\alpha }(\rho )$ reduces to Luo's $%
L(\rho )$ in Eq. (\ref{Luo-1}). Clearly, $Q_{\alpha }(\rho )\geq 0$. Note
that Tsallis' entropy is $S_{q}(\rho )=(1-\mbox{\rm Tr} \rho ^{q})/(q-1)$
indexed by also a parameter $q$ \cite{Tsalli}.

\section{Properties of $Q_{\protect\alpha }(\protect\rho )$}

Like WYD information, $Q_\alpha(\rho)$ inherits some interesting properties
from the WYD skew information. These properties are reminiscent of Tsallis
and Renyi entropies as generalized von Neumann entropies.

\begin{enumerate}
\item $Q_{\alpha }(\rho )$ is non-negative and it is always less than $n-1$,
i.e., $0\leq Q_{\alpha }(\rho )\leq n-1$, where $n$ is the dimensions of the
quantum system with system Hilbert space $C^{n}$.

\item For an $n$-dimensional completely mixed state $\rho =I/n$, von Neumann
entropy $S(\rho )=\ln n$. By the discussion in \cite{Luo2}, quantum
uncertainty of $\rho =I/n$\ should vanish. It is easy to verify that for the
completely mixed state $I/n$, the measure $Q_{\alpha }(\rho )$ vanishes.

\item It is not hard to know that $Q_{\alpha }(\rho )$ is convex because WYD
information is convex \cite{Lieb}. That is, $Q_{\alpha }(\sum_{i}\lambda
_{i}\rho _{i})\leq \sum_{i}\lambda _{i}Q_{\alpha }(\rho _{i})$, where $%
\lambda _{i}\geq 0$ and $\sum_{i}\lambda _{i}=1$.

\item The uncertainty measure $Q_{\alpha }(\rho )$ is always less than Luo's
one in Eq. (\ref{Luo-1}). It means that when $\alpha =1/2$, $Q_{\alpha
}(\rho )$ has the maximal value $L(\rho )$. That is,
\begin{equation}
Q_{\alpha }(\rho )\leq L(\rho ).  \label{ineq-1}
\end{equation}

The above inequality follows Eqs. (\ref{Luo-1}), (\ref{new-1}), and the
following inequality. $\lambda _{i}^{\alpha }\lambda _{j}^{1-\alpha
}+\lambda _{i}^{1-\alpha }\lambda _{j}^{\alpha }\geq 2\sqrt{\lambda
_{i}\lambda _{j}}$, for any $\alpha $, i.e., the arithmetic mean is greater
than the geometric mean, and the equality holds only when $\alpha =1/2$ or $%
\lambda _{1}=$ $\lambda _{2}=...=\lambda _{n}$ for any $\alpha .$

\item When $\alpha $ tends to $0$, $\lim Q_{\alpha }(\rho )=0$.
Symmetrically, when $\alpha $ tends to $1$, also $\lim Q_{\alpha }(\rho )=0$.

\item $Q_{\alpha }(\rho )$ is invariant under unitary transformations, i.e.,
$Q_{\alpha }(U\rho U^{\dagger })=Q_{\alpha }(\rho )$. This property follows
the definition in Eq. (\ref{new-1}) and that the eigenvalues of $\rho $ do
not vary under unitary transformations.

\item For pure states, von Neumann entropy $S(\rho )=0$. However, it can
also be argued that it is more intuitive if we require that all pure states
have the maximal quantum uncertainty \cite{Luo2}. In this sense, it is easy
to see that when $\rho $ is a pure state, $Q_{\alpha }(\rho )=n-1$ which is
maximal quantum uncertainty from Eq. (\ref{def-2}).

\item It is known that von Neumann entropy $S(\rho )$ is additive. That is, $%
S(\rho _{1}\otimes \rho _{2})=S(\rho _{1})+S(\rho _{2})$. Unfortunately, $%
Q_{\alpha }(\rho )$ is not additive. However, by the idea for the skew
information in \cite{Luo2} we can also show that $Q_{\alpha }(\rho )$ has
the following property. Let $P_{\alpha }(\rho )=Q_{\alpha }(\rho )/n$, $%
P_{\alpha }(\rho _{i})=Q_{\alpha }(\rho _{i})/\sqrt{n}$, where $Q_{\alpha
}(\rho _{i})=$ $\sqrt{n}-\mbox{\rm
Tr}\rho _{i}^{\alpha }\mbox{\rm Tr}\rho _{i}^{1-\alpha }$ by the Eq. (\ref%
{new-1}), $i=1$, $2$. From Eq. (\ref{new-1}), $Q_{\alpha }(\rho _{1}\otimes
\rho _{2})=n-\mbox{\rm Tr}\rho _{1}^{\alpha }\mbox{\rm Tr}\rho
_{1}^{1-\alpha }\mbox{\rm Tr}\rho _{2}^{\alpha }\mbox{\rm Tr}\rho
_{2}^{1-\alpha }$. Then we can derive

\begin{equation}
P_{\alpha }(\rho _{1}\otimes \rho _{2})+P_{\alpha }(\rho _{1})P_{\alpha
}(\rho _{2})=P_{\alpha }(\rho _{1})+P_{\alpha }(\rho _{2}).  \label{prob}
\end{equation}%
Luo derived Eq. (\ref{prob}) when $\alpha =1/2$ and\ thought that Eq. (\ref%
{prob}) with $\alpha =1/2$ resembles the probability law for union and
intersection of two events \cite{Luo2}.
\end{enumerate}

\section{The average of $Q_{\protect\alpha }(\protect\rho )$ as quantum
uncertainty}

If we wish to remove the dependence of $Q_{a}(\rho )$ on $\alpha $, we can
consider the average value of $Q_{\alpha }(\rho )$ over $\alpha $ as
follows. Let $Q^{\ast }(\rho )=\int_{0}^{1}Q_{\alpha }(\rho )d\alpha
=\sum_{i<k}(\lambda _{i}+\lambda _{k}-\int_{0}^{1}\lambda _{i}^{\alpha
}\lambda _{k}^{1-\alpha }d\alpha -\int_{0}^{1}\lambda _{i}^{1-\alpha
}\lambda _{k}^{\alpha }d\alpha )$. When $\lambda _{i}\lambda _{k}=0$, $%
\int_{0}^{1}\lambda _{i}^{\alpha }\lambda _{k}^{1-\alpha }d\alpha =0$. $%
\frac{{}}{{}}$When $\lambda _{i}=\lambda _{k}\neq 0$,$\ \int_{0}^{1}\lambda
_{i}^{\alpha }\lambda _{k}^{1-\alpha }d\alpha =\lambda _{i}$. Otherwise, $%
\int_{0}^{1}\lambda _{i}^{\alpha }\lambda _{k}^{1-\alpha }d\alpha =\frac{%
\lambda _{k}-\lambda _{i}}{\ln \lambda _{k}-\ln \lambda _{i}}$. Moreover, $%
\int_{0}^{1}\lambda _{i}^{1-\alpha }\lambda _{k}^{\alpha }d\alpha =\frac{%
\lambda _{k}-\lambda _{i}}{\ln \lambda _{k}-\ln \lambda _{i}}$. Let $\Delta
(\lambda _{i}$, $\lambda _{k})$ be defined by
\begin{equation}
\Delta (\lambda _{i},\lambda _{k})=\left\{
\begin{array}{rcc}
0 & : & \lambda _{i}\lambda _{k}=0\text{,} \\
2\lambda _{i} & : & \lambda _{i}=\lambda _{k}\neq 0, \\
\frac{2(\lambda _{k}-\lambda _{i})}{\ln \lambda _{k}-\ln \lambda _{i}} & : &
\text{otherwise.}%
\end{array}%
\right.
\end{equation}%
Then, $Q^{\ast }(\rho )=\sum_{i<k}[\lambda _{i}+\lambda _{k}-\Delta (\lambda
_{i},\lambda _{k})]$. By Eq. (\ref{def-2}), we can rewrite $Q^{\ast }(\rho
)=n-1-\sum_{i<k}\Delta (\lambda _{i},\lambda _{k})$.

Interestingly, $Q^{\ast }(\rho )$ has the following properties.

\begin{enumerate}
\item Clearly, $0\leq Q^{\ast }(\rho )\leq n-1$ because $0\leq Q_{\alpha
}(\rho )\leq n-1$.

\item $Q^{\ast }(\rho )$ is convex because $Q_{\alpha }(\rho )$ is convex.

\item $Q^{\ast }(\rho )\leq L(\rho )$. This follows Eq. (\ref{ineq-1}) and $%
\int_{0}^{1}Q_{\alpha }(\rho )d\alpha \leq \int_{0}^{1}L(\rho )d\alpha $.
The equality holds only when $\lambda _{1}=$ $\lambda _{2}=...=\lambda _{n}$
or $\alpha =1/2$.

\item For pure states, $Q^{\ast }(\rho )=n-1$, which is maximal quantum
uncertainty from the definition of $Q^{\ast }(\rho )$.

\item For an $n$-dimensional completely mixed state $\rho =I/n$, $Q^{\ast
}(\rho )=0$.

\item $Q^{\ast }(\rho )$\ is invariant under unitary transformations, i.e., $%
Q^{\ast }(U\rho U^{\dagger })=Q^{\ast }(\rho )$. \
\end{enumerate}

\begin{figure}[ht]
% open figure environment
\par
\begin{center}
% centre figure
\resizebox{0.6\columnwidth }{!}{     \includegraphics{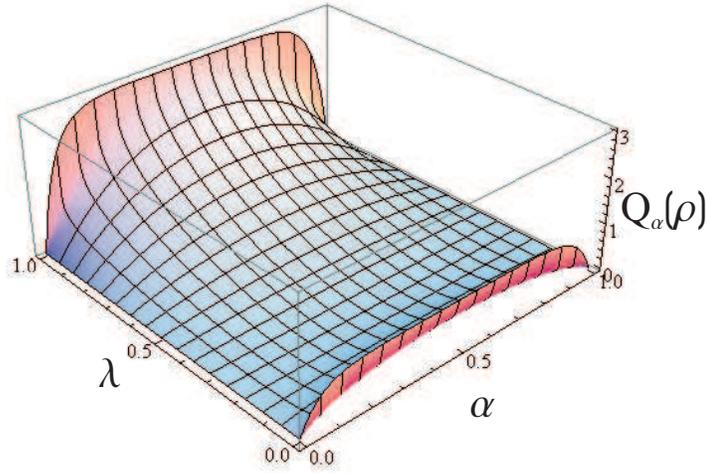}}
\end{center}
\caption{Wigner-Yanse-Dyson information for the Werner state as a function
of $\protect\alpha$ and $\protect\lambda$. At $\protect\lambda=1$, $Q_%
\protect\alpha(\protect\rho) =3$ regardless of the value of $\protect\alpha$
so there should be a straight-line (not shown) at that value.}
\label{fig:werner1}
\end{figure}

\begin{figure}[ht]
% open figure environment
\par
\begin{center}
% centre figure
\resizebox{0.6\columnwidth }{!}{     \includegraphics{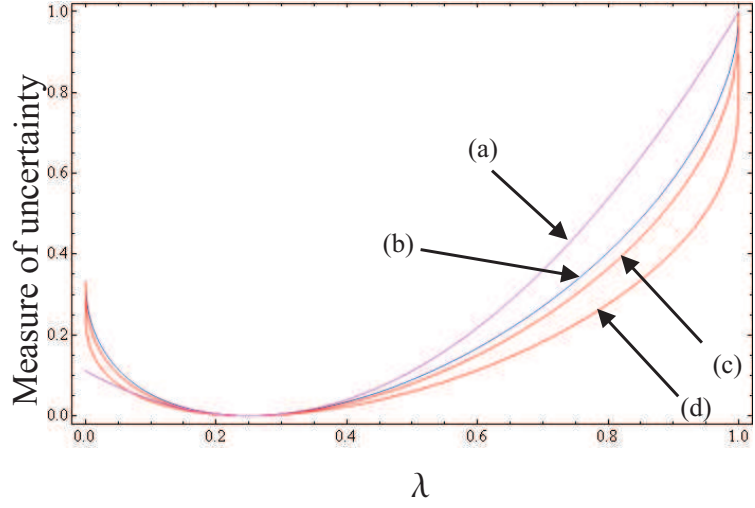}}
\end{center}
\caption{Different measures (normalized) to unity for the pure state
for (a) Brukner-Zeilinger information, (b) Luo information (c)
Wigner-Yanase-Dyson
(WYD) for $\protect\alpha =1/3$, i.e. $Q_{1/3}(\protect\rho)$, and (d) $%
Q^\ast(\protect\rho)$.} \label{fig:werner2}
\end{figure}

\begin{figure}[ht]
% open figure environment
\par
\begin{center}
% centre figure
\resizebox{0.6\columnwidth }{!}{     \includegraphics{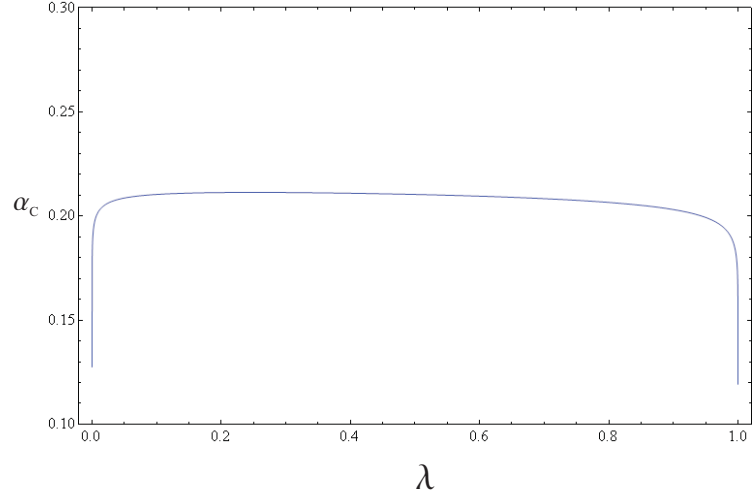}}
\end{center}
\caption{Critical values of $\protect\alpha$ as a function of the
state parameter $\protect\lambda$} \label{fig:werner3}
\end{figure}

Next we consider the Werner state $\displaystyle\rho =\frac{4\lambda -1}{3}%
|\Psi ^{-}\rangle \langle \Psi ^{-}|+\frac{(1-\lambda )}{3}\frac{I}{4}$
where $\displaystyle|\Psi ^{-}\rangle =\frac{1}{\sqrt{2}}(|01\rangle
-|10\rangle )$ is the singlet state for two qubits. Fig. \ref{fig:werner1}
shows the Wigner-Yanase-Dyson (WYD) information for the Werner state as a
function of the parameters $\alpha $ and $\lambda $. Clearly, WYD
information is symmetric with respect to $\alpha $ and acquires its maximum
value at $\alpha =1/2$ (Luo's value). In Fig. \ref{fig:werner2}, we plot
various measures of information as a function of the state parameter $%
\lambda $. In (a), we consider Brukner Zeilinger (normalized) measure
defined by $\displaystyle I_{BZ}=\frac{n}{n-1}(\mbox{\rm Tr}\rho ^{2}-1/n)$
with $n=4$ in the example. We consider $\displaystyle\frac{1}{n-1}Q_{\alpha
}(\rho ))$ for (b) $\alpha =1/2$ (Luo information) and (c) $\alpha =1/3$,
and in (d) we evaluate $Q^{\ast }(\rho )$. Note that the minimal value of
zero is obtained for the maximally mixed state, i.e. when $\lambda =1/4$. It
is also interesting to note that for each $\lambda $, if one computes the
critical value of $\alpha =\alpha _{c}$ such that $Q_{\alpha _{c}}(\rho
(\lambda ))=Q^{\ast }(\rho (\lambda ))$, such a function is a slowly varying
function of $\lambda $. The plot of $\alpha -c$ as a function of $\lambda $
is shown in Fig. \ref{fig:werner3}.

Incidentally let us consider Hansen's example in \cite{Hansen} where he
considered $\rho _{12}^{\ast }=\left(
\begin{tabular}{llll}
$7$ & $5$ & $5$ & $6$ \\
$5$ & $6$ & $2$ & $5$ \\
$5$ & $2$ & $6$ & $5$ \\
$6$ & $5$ & $5$ & $7$%
\end{tabular}%
\right) $. Note that $\rho _{12}^{\ast }$ is not a density operator because $%
tr(\rho _{12}^{\ast })\neq 1$. We let $\rho _{12}=\rho _{12}^{\ast }/26$.
Thus, $\rho _{12}$\ becomes a density operator. By calculating, von Neumann
entropy $S(\rho _{12})=\allowbreak \allowbreak 0.603\,19$, Luo's quantum
uncertainty $L(\rho _{12})=\allowbreak 1.\,\allowbreak 538\,5$, our quantum
uncertainty $Q_{1/4}(\rho _{12})=1.2213$ and $Q^{\ast }(\rho
_{12})=\allowbreak 1.\,\allowbreak 0748$.

In summary, by averaging Wigner-Yanase-Dyson information we derive the
measure $Q_{\alpha }(\rho )$ indexed by $0<\alpha <1$\ of quantum
uncertainty for a mixed state $\rho $. We demonstrate the interesting
properties of $Q_{\alpha }(\rho )$. The result is reminiscent of the
extension to generalized entropies for the von Neumann entropy. To remove
the dependence on the parameter $\alpha $, we can take the average $Q^{\ast
}(\rho )$ of $Q_{\alpha }(\rho )$ over $\alpha $ and derive a measure of
quantum uncertainty of a mixed state. Finally we study some of the
properties of $Q^{\ast }(\rho )$.

\section{Appendix A Proof of the invariance under unitary transformations}

\setcounter{equation}{0} \renewcommand{\theequation}{A\arabic{equation}} \ \

(A). The proof of $I_{\alpha }(U\rho U^{\dagger }$, $X)=I_{\alpha }(\rho $, $%
U^{\dagger }XU)$

By the definition, $I_{\alpha }(U\rho U^{\dagger }$, $X)=\mbox{\rm Tr}
(U\rho U^{\dagger }X^{2})-\mbox{\rm Tr} ((U\rho U^{\dagger })^{\alpha
}X(U\rho U^{\dagger })^{1-\alpha }X)$ and $I_{\alpha }(\rho $, $U^{\dagger
}XU)=\mbox{\rm Tr} (\rho (U^{\dagger }XU)^{2})-\mbox{\rm Tr} (\rho ^{\alpha
}(U^{\dagger }XU)\rho ^{1-\alpha }(U^{\dagger }XU))$. By calculating,
\begin{equation}
\mbox{\rm Tr} (\rho (U^{\dagger }XU)^{2})=\mbox{\rm Tr} (\rho (U^{\dagger
}XU)(U^{\dagger }XU))=\mbox{\rm Tr} (U\rho U^{\dagger }X^{2}).  \label{C-1}
\end{equation}%
It is easy to see that $U\rho U^{\dagger }$ is self-adjoint. Let $\rho $
have a spectral representation

\begin{equation}
\rho =\lambda _{1}|x_{1}\rangle \langle x_{1}|+....+\lambda
_{n}|x_{n}\rangle \langle x_{n}|.  \label{spectral-1}
\end{equation}
Then, we obtain the following spectral representation of $U\rho U^{\dagger }$%
. $U\rho U^{\dagger }=\lambda _{1}U|x_{1}\rangle \langle x_{1}|U^{\dagger
}+....+\lambda _{n}U|x_{n}\rangle \langle x_{n}|U^{\dagger }$. Note that
orthonormal basis $\{Ux_{1}$,..., $Ux_{n}\}$\ consists of eigenvectors of $%
U\rho U^{\dagger }$ and $\lambda _{1}$, ..., $\lambda _{n}$ are the
corresponding eigenvalues. Thus,
\begin{equation}
(U\rho U^{\dagger })^{\alpha }=\lambda _{1}^{\alpha }U|x_{1}\rangle \langle
x_{1}|U^{\dagger }+....+\lambda _{n}^{\alpha }U|x_{n}\rangle \langle
x_{n}|U^{\dagger }=U\rho ^{\alpha }U^{\dagger }\text{.}  \label{C-3}
\end{equation}%
As well,
\begin{equation}
(U\rho U^{\dagger })^{1-\alpha }=U\rho ^{1-\alpha }U^{\dagger }.  \label{C-4}
\end{equation}%
It is ready to get the following from Eqs. (\ref{C-3}) and (\ref{C-4}).

\begin{equation}
\mbox{\rm Tr} ((U\rho U^{\dagger })^{\alpha }X(U\rho U^{\dagger })^{1-\alpha
}X)=\mbox{\rm Tr} (U\rho ^{\alpha }U^{\dagger }XU\rho ^{1-\alpha }U^{\dagger
}X)=\mbox{\rm Tr} (\rho ^{\alpha }(U^{\dagger }XU)\rho ^{1-\alpha
}(U^{\dagger }XU)).  \label{C-2}
\end{equation}%
From Eqs. (\ref{C-1}) and (\ref{C-2}), we finish this proof.

(B). The proof of $I_{\alpha }(U\rho U^{\dagger }$, $UXU^{\dagger
})=I_{\alpha }(\rho $, $X)$

It is straightforward to get the proof from Eqs. (\ref{C-3}) and (\ref{C-4}).

\section{Acknowledgment}

The paper was supported by NSFC(Grants No.10875061, 60673034). The work was
done partially while the first author was visiting the Institute for
Mathematical Sciences, National University of Singapore in August, 2008. KLC
would like to acknowledge financial support by the National Research
Foundation \& Ministry of Education, Singapore, for his visit and
collaboration at Tsinghua University.

\end{document}